\documentclass[12pt]{article}
\usepackage{type1cm}
\usepackage[dvipdfmx]{graphicx}
\usepackage{cite}

\usepackage{amsmath,amssymb,bm,amsthm,cases}

\def\beq{\begin{equation}}
\def\eeq{\end{equation}}
\def\nbeq{\begin{equation*}}
\def\neeq{\end{equation*}}

\def\<{\langle}
\def\>{\rangle}

\begin{document}
\title{Extensive increase of entropy in quantum quench}
\author{Takashi Mori \\
{\it
Department of Physics, Graduate School of Science,} \\
{\it The University of Tokyo, Bunkyo-ku, Tokyo 113-0033, Japan
}}
\date{}
\maketitle
\begin{abstract}
In the setup of isolated quantum systems, it is proved that the thermodynamic entropy and the diagonal entropy must increase extensively in any nontrivial quantum quench.
The extensive increase of the thermodynamic entropy is shown for any initial state (even for a pure state) that represents thermal equilibrium.
On the other hand, the extensive increase of the diagonal entropy is shown for any stationary initial state under the condition that both the pre-quench and the post-quench Hamiltonians satisfy the eigenstate thermalization hypothesis.
\end{abstract}

\section{Introduction}
\label{sec:intro}

Microscopic understanding of the second law of thermodynamics and irreversibility is one of the fundamental problems in theoretical physics.
According to different setups, several microscopic derivations of the second law type inequalities have been presented~\cite{Klein1931,Pusz1978,Lenard1978,Jarzynski1997,Jarzynski2004,Campisi2008,Polkovnikov2011, Sagawa_lecture2013,Ikeda2015,Tasaki2016,Tajima-Wakakuwa_arXiv2016}.
We focus on isolated quantum systems, in which the density matrix $\rho$ is given as the microscopic description of the state of the system and it evolves in time according to the Liouville-von-Neumann equation.
The second law of thermodynamics implies that the entropy increases or remains constant under an adiabatic process.
Here, the adiabatic process means the time evolution under the time-dependent Hamiltonian.
In the context of quantum mechanics, the terminology of ``adiabatic'' is sometimes used only for the case in which the change of the Hamiltonian is infinitely slow, but in this paper, the process under an arbitrarily fast change of the Hamiltonian is said to be adiabatic as long as the system is isolated from the environment and the heat does not flow between them.
Because the von Neumann entropy is invariant under the unitary dynamics, we must consider different definitions of the entropy to show the second law\footnote
{We can also formulate the second law in terms of the mechanical quantities only, e.g., the energy and the work, instead of considering the entropy~\cite{Pusz1978,Lenard1978,Jarzynski1997,Jarzynski2004,Tasaki2016}.}.

So far, for some appropriately defined entropies, it has been shown that $S_f-S_i\geq 0$, where $S_i$ and $S_f$ are the entropies before and after an adiabatic process, respectively~\cite{Klein1931,Campisi2008,Polkovnikov2011,Ikeda2015,Tajima-Wakakuwa_arXiv2016}.

However, the inequality $S_f-S_i\geq 0$ only tells us that the entropy does not decrease in an adiabatic process.
If we consider a non-quasi-static adiabatic process in which the Hamiltonian changes rapidly, it is expected that the amount of increase of the entropy is extensively large, $S_f-S_i=\mathcal{O}(N)$, where $N$ characterizes the system size such as the number of particles, but no general microscopic proof of this stronger statement has been provided.

It is desired to prove $S_f-S_i=\mathcal{O}(N)$ microscopically, which we call the ``second law with strict irreversibility''.
In this paper, we consider this problem for quantum quench, in which the Hamiltonian changes instantaneously at a certain time from, say, $H_i$ to $H_f$.
We will show the second law with strict irreversibility for two different setups, in which the thermodynamic entropy~\cite{Tasaki2016} and the diagonal entropy~\cite{Klein1931,Polkovnikov2011} are considered, respectively.

This paper is organized as follows.
In Sec.~\ref{sec:entropy}, the three definitions of the entropy are summarized.
In Sec.~\ref{sec:previous}, previous standard derivations of the second law of thermodynamics are presented, and in Sec.~\ref{sec:problem}, the problem in the previous derivations of the second law is discussed.
In Sec.~\ref{sec:inequality}, a useful inequality on the relative entropy is presented, and by using this inequality, we prove the second law with strict irreversibility in Sec.~\ref{sec:main}, which is the main part of this paper.
Section~\ref{sec:summary} is devoted to the summary and the discussion.

\section{Three definitions of the entropy}
\label{sec:entropy}

In this paper, we consider the second law for the thermodynamic entropy and the diagonal entropy.
In this section, the definitions of these entropies as well as more familiar von Neumann entropy and the relation among them are summarized for completeness.
Essentially the same content of this section is found in Supplementary Material of Ref.~\cite{Tasaki2016}.

For a given density matrix $\rho$, the von Neumann entropy is defined as
\beq
S(\rho):=-\mathrm{Tr}\rho\ln\rho.
\eeq
This gives the correct equilibrium entropy if $\rho$ is chosen as the Gibbs state, $\rho=e^{-\beta H}/\mathrm{Tr}e^{-\beta H}$ (or the microcanonical density matrix), with $\beta$ the inverse temperature and $H$ the Hamiltonian.
However, it is known that the von Neumann entropy is invariant under any unitary time evolution, and hence it fails to show any increase in an adiabatic process.
It means that the von Neumann entropy is not appropriate when we discuss the change of the entropy in an adiabatic process.

As the second definition, we introduce the so-called diagonal entropy $S_{\mathrm{D}}(\rho)$~\cite{Klein1931,Polkovnikov2011}.
The diagonal entropy is defined as the von Neumann entropy of the ``diagonal ensemble'',
\beq
S_{\mathrm{D}}(\rho):=S(\rho_{\mathrm{D}})=-\mathrm{Tr}\rho_{\mathrm{D}}\ln\rho_{\mathrm{D}},
\eeq
where $\rho_{\mathrm{D}}$ is defined as the diagonal part of $\rho$ in the basis of energy eigenstates.
It is known that the diagonal entropy does change in an adiabatic process, and as we will discuss in the next section, the diagonal entropy does not decrease for an adiabatic process under certain conditions.

The third definition is referred to as the thermodynamic entropy~\cite{Tasaki2016}, which is defined as the von Neumann entropy of the Gibbs state that has the expectation value of the energy identical to that of the actual state $\rho$,
\beq
S_{\mathrm{TD}}(\rho):=S(\rho_{\mathrm{G}}),
\eeq
where $\rho_{\mathrm{G}}=e^{-\beta H}/\mathrm{Tr}e^{-\beta H}$ with $\beta$ determined by the equality $\mathrm{Tr}\rho H=\mathrm{Tr}\rho_{\mathrm{G}}H$.
This can also show the time dependence.

The difference among the three definitions of the entropy is interpreted as the different assignment of the microscopic states.
In the von Neumann entropy, the full information of the microscopic state $\rho$ is kept, while in the diagonal entropy, only the information of the diagonal elements is kept, which means that we regard the two microscopic states $\rho^{(1)}$ and $\rho^{(2)}$ as the ``identical state'' if all the diagonal elements of these density matrices are identical, $\rho_{\mathrm{D}}^{(1)}=\rho_{\mathrm{D}}^{(2)}$.
In the thermodynamic entropy, all the information of the state $\rho$ except for the internal energy $\mathrm{Tr}\rho H$ is lost.
This means that we distinguish the two states $\rho^{(1)}$ and $\rho^{(2)}$ only by the values of the internal energy.
Thus, $S(\rho)$ is the most ``fine-grained'' entropy and $S_{\mathrm{TD}}(\rho)$ is the most ``coarse-grained'' entropy.
Accordingly, the following inequality holds~\cite{Tasaki2016}:
\beq
S(\rho)\leq S_{\mathrm{D}}(\rho)\leq S_{\mathrm{TD}}(\rho).
\label{eq:relation}
\eeq

\section{Previous derivations of the second law of thermodynamics}
\label{sec:previous}

According to different definitions of the entropy, different assumptions on the initial state, and different setups (whether the system is isolated~\cite{Klein1931,Lenard1978,Jarzynski1997} or is in contact with a thermal reservoir~\cite{Jarzynski2004,Sagawa_lecture2013,Iyoda2016}, and whether an adiabatic process is modeled by a time-dependent Hamiltonian $H(t)$ or considering a time-independent Hamiltonian including the apparatus operating the system~\cite{Tasaki2016}, and so on), there are various kinds of the second-law type inequalities.
In this section, the two standard microscopic derivations of the second law are explained.

We consider an isolated quantum system with the initial state $\rho^{(i)}$.
An adiabatic process is described by a time-dependent Hamiltonian $H(t)$ ($0\leq t\leq\tau$) with $H(0)=H_i$ and $H(\tau)=H_f$.
Throughout the paper, we set $\hbar=1$.
The time evolution operator is then given by $U_{\tau}=\mathcal{T}e^{-i\int_0^{\tau}H(t)dt}$, where $\mathcal{T}$ stands for the time-ordering operator.
Then, the final state is given by $\rho^{(f)}=U_{\tau}\rho^{(i)}U_{\tau}^{\dagger}$.

The first setup is summarized as follows:
\begin{description}
\item[(I)] The initial state is given by the Gibbs state $\rho^{(i)}=\rho^{(i)}_{\mathrm{G}}=e^{-\beta H_i}/\mathrm{Tr}e^{-\beta H_i}$.
Compare the thermodynamic entropies before and after the adiabatic process.
\end{description}
Since the initial state is given by the Gibbs state, the thermodynamic entropy in the initial state is written as $S_{\mathrm{TD}}(\rho^{(i)})=S(\rho^{(i)})$.
Because the von Neumann entropy is invariant under unitary time evolution, we have $S(\rho^{(i)})=S(\rho^{(f)})$.
By using the relation (\ref{eq:relation}), $S(\rho^{(f)})\leq S_{\mathrm{TD}}(\rho^{(f)})$, and thus we obtain
\beq
S_{\mathrm{TD}}(\rho^{(i)})\leq S_{\mathrm{TD}}(\rho^{(f)}).
\label{eq:2ndI}
\eeq
This is the statement of the second law of thermodynamics.

The second setup is given as follows:
\begin{description}
\item[(II)] The initial state is given by a stationary state $\rho^{(i)}=\rho^{(i)}_{\mathrm{D}}$, i.e., the initial density matrix $\rho^{(i)}$ is diagonal in the basis of the eigenstates of $H^{(i)}$.
Compare the diagonal entropies before and after the adiabatic process.
\end{description}
From the assumption, the diagonal entropy of the initial state is $S_{\mathrm{D}}(\rho^{(i)})=S(\rho^{(i)})$.
By using the invariance of the von Neumann entropy, $S(\rho^{(i)})=S(\rho^{(f)})$.
The relation (\ref{eq:relation}) then leads to $S(\rho^{(f)})\leq S_{\mathrm{D}}(\rho^{(f)})$.
Thus we conclude
\beq
S_{\mathrm{D}}(\rho^{(i)})\leq S_{\mathrm{D}}(\rho^{(f)}),
\label{eq:2ndII}
\eeq
which is the second-law type inequality for the diagonal entropy.

Here, it is remarked that the assumption of the stationary initial state made in (II) is not necessary under certain conditions, see Ref.~\cite{Ikeda2015}.
The idea is that we introduce the waiting time $t_w$ before the adiabatic process.
The initial density matrix is then replaced by $\rho^{(i)}(t_w)=e^{-iH_it_w}\rho^{(i)}e^{iH_it_w}$.
It is shown that for almost all large enough waiting times $t_w$, the diagonal entropy does not decrease, $S_{\mathrm{D}}(\rho^{(i)})\leq S_{\mathrm{D}}(\rho^{(f)})$ even if $\rho^{(i)}\neq\rho^{(i)}_{\mathrm{D}}$~\cite{Ikeda2015}.

\section{Problem to be solved}
\label{sec:problem}

Although the microscopic derivations of the second law presented in the previous section are simple and clear, there are remaining fundamental problems to be solved.

In this paper, we consider the following problem.
As mentioned in Introduction, the inequality $S^{(i)}\leq S^{(f)}$ only insists that the entropy does not decrease in an adiabatic process, but it does not tell us about what amount of entropy increases.
Although it is expected that the entropy increases extensively in a non-quasi-static adiabatic process, that is, $S^{(f)}-S^{(i)}=\mathcal{O}(N)>0$, it has not been shown in a general setup.
It is desired to prove the strict irreversibility in order to gain insights into more satisfactory microscopic understanding of the second law and the irreversibility.

In this paper, this problem is solved for the limiting case, that is, in quantum quench, in which the Hamiltonian is suddenly switched from $H_i$ to $H_f$ at time $t=0$.
For $t>0$, the Hamiltonian is fixed to be $H(t)=H_f$.
We shall prove the strict irreversibility in this setup.

\section{Inequality for the relative entropy}
\label{sec:inequality}

In order to derive the strict irreversibility in quantum quench, we should prepare some sophisticated theoretical tool.
In Sec.~\ref{sec:previous}, two standard derivations of the second law are demonstrated, in which the invariance of the von Neumann entropy under the unitary dynamics and the relation (\ref{eq:relation}) are used.

On the other hand, the second-law type inequalities are also derived by using the non-negativity of the relative entropy in many cases~\cite{Sagawa_lecture2013}.
The relative entropy of a density matrix $\sigma$ with respect to another density matrix $\rho$ is defined as
\beq
S(\sigma\|\rho):=\mathrm{Tr}\sigma(\ln\sigma-\ln\rho).
\label{eq:relative}
\eeq
It is shown that $S(\sigma\|\rho)\geq 0$ for any $\sigma$ and $\rho$, which is called the non-negativity of the relative entropy.
From this property, we can provide another proof of (\ref{eq:2ndI}) and (\ref{eq:2ndII}).

Recently, in the study on the ensemble equivalence of two general states $\sigma$ and $\rho$, a useful inequality on the relative entropy is derived~\cite{Mori2016_macrostate}.
In order to explain it, we introduce the concept of macrovariables, which have vanishingly small fluctuations and obey the large-deviation principle in an equilibrium state.
We do not give the precise definition of macrovariables, but one can suppose $X$ written as $X=(1/N)\sum_{i=1}^NO_i$, where $i$ denotes each particle (or each lattice site) and $O_i$ is a local operator acting to the particles (or the lattice sites) close to $i$, as a macrovariable.
Hamiltonian per particle $H/N$ and the magnetization density in a spin system are representative examples of macrovariables.

We consider the sequence of the density matrices $\{\rho_N\}_{N\in\mathbb{N}}$ and $\{\sigma_N\}_{N\in\mathbb{N}}$, which specify the way of taking the thermodynamic limit $N\rightarrow+\infty$.
The states specified by the thermodynamic limit of $\rho_N$ and $\sigma_N$ are simply denoted by $\rho$ and $\sigma$, respectively.

For a macrovariable $X$, we introduce the large-deviation rate function as follows:
\beq
I_{\rho}(x):=-\limsup_{N\rightarrow\infty}\frac{1}{N}\ln\mathrm{Prob}_{\rho_N}\left[X\in[x,x+dx)\right],
\eeq
and $I_{\sigma}(x)$ is similarly defined.
Here,
\beq
\mathrm{Prob}_{\rho_N}\left[X\in[x,x+dx)\right]:=\textrm{Tr}\rho_N\mathsf{P}_{X\in[x,x+dx)},
\eeq
where $\mathsf{P}_{X\in[x,x+dx)}$ is the projection operator onto the Hilbert subspace spanned by $\{|\psi\>: X|\psi\>=x'|\psi\>, x'\in[x,x+dx)\}$.

The set of typical values of $X$ in the state $\rho$($\sigma$) is denoted by $\mathcal{E}_{\rho(\sigma)}:=\{x:I_{\rho(\sigma)}(x)=0\}$.
For a macrovariable, whose fluctuation is vanishingly small in the thermodynamic limit, $N\rightarrow\infty$, it is expected that there is only single typical value, $\mathcal{E}_{\rho}=\{x_{\rho}^*\}$ and $\mathcal{E}_{\sigma}=\{x_{\sigma}^*\}$.
In this section, we assume it\footnote
{We do not consider the case of phase coexistence, where the rate function has some flat region of $I_{\rho}(x)=0$.}.

When $\rho$ is given by the Gibbs state $\rho=e^{-\beta H}/\mathrm{Tr}e^{-\beta H}$, it is rigorously proven that $\mathcal{E}_{\rho}$ actually has a unique element for any $\beta >0$ in one-dimensional spin systems with translation invariance~\cite{Ogata2010}, and for $\beta<\beta^*$ with some $\beta^*>0$ in higher dimensional spin systems with translation invariance~\cite{Netocny2004,Lenci2005}.

In Ref.~\cite{Mori2016_macrostate}, it was shown that the following inequality is satisfied for all $\alpha>1$:
\beq
I_{\sigma}(x)\geq\frac{\alpha-1}{\alpha}\left[I_{\rho}(x)-s_{\alpha}(\sigma\|\rho)\right],
\label{eq:inequality}
\eeq
where $s_{\alpha}(\sigma\|\rho)$ is the specific relative Renyi entropy defined as
\beq
s_{\alpha}(\sigma\|\rho):=\lim_{N\rightarrow\infty}\frac{S_{\alpha}(\sigma_N\|\rho_N)}{N},
\eeq
and the relative Renyi entropy is defined as
\beq
S_{\alpha}(\sigma\|\rho):=\frac{1}{\alpha-1}\ln\mathrm{Tr}\sigma^{\alpha}\rho^{1-\alpha}.
\eeq
It is known that $S_{\alpha}(\sigma\|\rho)\rightarrow S(\sigma\|\rho)$ as $\alpha\rightarrow 1$.

Let us substitute $x=x_{\sigma}^*$ into (\ref{eq:inequality}).
By definition $I_{\sigma}(x_{\sigma}^*)=0$, and hence
\beq
s_{\alpha}(\sigma\|\rho)\geq I_{\rho}(x_{\sigma}^*).
\eeq
By taking the limit of $\alpha\rightarrow 1$, for sufficiently large system size $N$, we have\footnote
{Here we assumed the continuity of $s_{\alpha}(\sigma\|\rho)$ at $\alpha=1$ as a function of $\alpha$.
Since it is known that $s_{\alpha}(\sigma\|\rho)$ is a convex function of $\alpha$ for $1\leq\alpha\leq 2$~\cite{Ando1979}, the assumption of continuity is quite natural.
If $s_{\alpha}(\sigma\|\rho)$ is discontinuous at $\alpha=1$, it implies $s_{\alpha}(\sigma\|\rho)=+\infty$ for all $\alpha>1$, which is unlikely.
If we can check $s_2(\sigma\|\rho)<+\infty$, we can conclude the continuity at $\alpha=1$.}
\beq
S(\sigma_N\|\rho_N)\geq NI_{\rho}(x_{\sigma}^*)+o(N).
\label{eq:ineq_relative}
\eeq
If $\sigma$ and $\rho$ are macroscopically distinct states, there should be some macrovariable $X$ such that $x_{\sigma}^*\neq x_{\rho}^*$, and hence $I_{\rho}(x_{\sigma}^*)>0$.
The inequality (\ref{eq:ineq_relative}) implies that the relative entropy of two macroscopically distinct states must be extensively large.

This inequality is much stronger than the non-negativity of the relative entropy $S(\sigma\|\rho)\geq 0$, and it provides a powerful tool to derive the second law of thermodynamics with strict irreversibility.

In the remaining part of the paper, we shall drop the subscript $N$ of $\rho_N$ and $\sigma_N$ because there would be no confusion.
We always consider a sufficiently large system size $N$.

\section{Strict irreversibility in quantum quench}
\label{sec:main}

In quantum quench, the Hamiltonian changes suddenly from $H_i$ to $H_f$.
The eigenstates and the eigenvalues of $H_a$ ($a=i$ or $f$) are denoted by $|\phi_n^{(a)}\>$ and $E_n^{(a)}$ with $H_{a}|\phi_n^{(a)}\>=E_n^{(a)}|\phi_n^{(a)}\>$.
Since the post-quench Hamiltonian is fixed to be $H_f$, the time evolution operator is given by $U_{\tau}=e^{-iH_f\tau}$.
We consider the same setup as (I) and (II) in Sec.~\ref{sec:previous}.

\subsection{Setup of (I)}

In the first setup (I), the initial state is assumed to be a Gibbs state $\rho^{(i)}=\rho^{(i)}_{\mathrm{G}}=e^{-\beta H_i}/\mathrm{Tr}e^{-\beta H_i}$ (but later it turns out that this assumption can be removed; see below).
After a quantum quench, the final state is given by $\rho^{(f)}=e^{-iH_f\tau}\rho_{\mathrm{G}}^{(i)}e^{iH_f\tau}$.
The corresponding Gibbs state is given by $\rho^{(f)}_{\mathrm{G}}=e^{-\beta_fH_f}/\mathrm{Tr}e^{-\beta_fH_f}$, where $\beta_f$ is determined by
\beq
\mathrm{Tr}\rho^{(f)}H_f=\mathrm{Tr}\rho^{(f)}_{\mathrm{G}}H_f.
\label{eq:betaI}
\eeq
Since $\mathrm{Tr}\rho^{(f)}H_f=\mathrm{Tr}\rho^{(i)}_{\mathrm{G}}H_f$, $\beta_f$ satisfies
\beq
\mathrm{Tr}\rho^{(i)}_{\mathrm{G}}H_f=\mathrm{Tr}\rho^{(f)}_{\mathrm{G}}H_f.
\label{eq:betaI2}
\eeq

The relative entropy $S(\rho^{(i)}_{\mathrm{G}}\|\rho^{(f)}_{\mathrm{G}})$ is calculated as
\beq
S(\rho^{(i)}_{\mathrm{G}}\|\rho^{(f)}_{\mathrm{G}})=S(\rho^{(f)}_{\mathrm{G}})-S(\rho^{(i)}_{\mathrm{G}})+\beta_f \left[\mathrm{Tr}\rho^{(i)}_{\mathrm{G}}H_f-\mathrm{Tr}\rho^{(f)}_{\mathrm{G}}H_f\right].
\label{eq:relative_eq}
\eeq
By using (\ref{eq:betaI2}), we have
\beq
S(\rho^{(i)}_{\mathrm{G}}\|\rho^{(f)}_{\mathrm{G}})=S(\rho^{(f)}_{\mathrm{G}})-S(\rho^{(i)}_{\mathrm{G}})=S_{\mathrm{TD}}(\rho^{(f)})-S_{\mathrm{TD}}(\rho^{(i)}),
\eeq
i.e., the relative entropy is nothing but the change of the thermodynamic entropy.
By using the non-negativity of the relative entropy, we obtain the usual second-law inequality $S_{\mathrm{TD}}(\rho^{(f)})-S_{\mathrm{TD}}(\rho^{(i)})\geq 0$.
We can say more by using (\ref{eq:ineq_relative}):
\beq
S_{\mathrm{TD}}(\rho^{(f)})-S_{\mathrm{TD}}(\rho^{(i)})\geq NI_f(x_i^*)+o(N),
\label{eq:TD_inequality}
\eeq
where $x_i^*$ is the typical value of $X$ in the state $\rho^{(i)}_{\mathrm{G}}$, and we have used the shorthand notation $I_f(x)=I_{\rho^{(f)}_{\mathrm{G}}}(x)$.
It should be noted that the inequality~(\ref{eq:TD_inequality}) is satisfied for any choice of $X$.
Different choices of the macrovariable $X$ will give rise to different lower bounds on the change of the thermodynamic entropy.

In any nontrivial quantum quench, the initial equilibrium state $\rho^{(i)}_{\mathrm{G}}$ and the final equilibrium state $\rho^{(f)}_{\mathrm{G}}$ are distinct, which implies that there exists some macrovariable $X$ such that $x_i^*\neq x_f^*$ and therefore $I_f(x_i^*)>0$ (and independent of $N$), where $x_f^*$ is the typical value of $X$ in the state $\rho^{(f)}_{\mathrm{G}}$.
Therefore,
\beq
S_{\mathrm{TD}}(\rho^{(f)})-S_{\mathrm{TD}}(\rho^{(i)})=\mathcal{O}(N).
\label{eq:2nd_law_I}
\eeq
If $S_{\mathrm{TD}}(\rho^{(f)})-S_{\mathrm{TD}}(\rho^{(i)})=o(N)$, it implies that $\rho^{(i)}_{\mathrm{G}}$ and $\rho^{(f)}_{\mathrm{G}}$ are macroscopically indistinguishable.
Equation (\ref{eq:2nd_law_I}) is the desired second law with strict irreversibility for the thermodynamic entropy.

It should be remarked that, up to (\ref{eq:relative_eq}), we do not use the assumption of $\rho^{(i)}=\rho^{(i)}_{\mathrm{G}}$.
For a general initial state $\rho^{(i)}$, $\beta_f$ is determined by (\ref{eq:betaI}), which is also written as
\beq
\mathrm{Tr}\rho^{(i)}H_f=\mathrm{Tr}\rho^{(f)}_{\mathrm{G}}H_f,
\eeq
and hence, we have
\beq
S(\rho^{(i)}_{\mathrm{G}}\|\rho^{(f)}_{\mathrm{G}})=S(\rho^{(f)}_{\mathrm{G}})-S(\rho^{(i)}_{\mathrm{G}})+\beta_f\left[\mathrm{Tr}\rho^{(i)}_{\mathrm{G}}H_f-\mathrm{Tr}\rho^{(i)}H_f\right].
\eeq
Even if $\rho^{(i)}\neq\rho^{(i)}_{\mathrm{G}}$, (\ref{eq:2nd_law_I}) is obtained as long as $\mathrm{Tr}\rho^{(i)}H_f=\mathrm{Tr}\rho^{(i)}_{\mathrm{G}}H_f$.

In the studies on thermalization in isolated quantum systems, it is pointed out that a ``typical state'' $\rho^{(i)}$ represents the thermal equilibrium in the sense that $\mathrm{Tr}\rho^{(i)}O=\mathrm{Tr}\rho^{(i)}_{\mathrm{G}}O$ for any local operator $O$~\cite{Neumann1929,Tasaki1998,Goldstein2006,Reimann2007}.
As long as the initial state represents the equilibrium state in this sense, we can conclude the second law with strict irreversibility in quantum quench.

Along a similar discussion of Ref.~\cite{Ikeda2015}, let us consider some waiting time $t_w$ before the adiabatic process, by which the initial state is replaced by $\rho^{(i)}(t_w)=e^{-iH_it_w}\rho^{(i)}e^{iH_it_w}$.
Under some assumptions such as the eigenstate thermalization hypothesis~\cite{Deutsch1991,Srednicki1994,Rigol2008} or a moderate energy distribution of the initial state~\cite{Tasaki1998,Reimann2007}, it is shown that the system reaches thermal equilibrium in the sense that
\beq
\mathrm{Tr}\rho^{(i)}(t_w)O\approx\mathrm{Tr}\rho^{(i)}_{\mathrm{G}}O
\eeq
for almost all large enough $t_w$ and for any local operator $O$, where the difference between the left-hand side and the right-hand side is exponentially small in the system size $N$.
By putting $O=H_f$, we can conclude that $\mathrm{Tr}\rho^{(i)}(t_w)H_f\approx\mathrm{Tr}\rho^{(i)}_{\mathrm{G}}H_f$, and hence
\beq
S_{\mathrm{TD}}(\rho^{(f)})-S_{\mathrm{TD}}(\rho^{(i)})=\mathcal{O}(N)
\eeq
for almost all choice of $t_w$ (here, $\rho^{(f)}=e^{-iH_f\tau}\rho^{(i)}(t_w)e^{iH_f\tau}$).
Thus, the assumption of $\rho^{(i)}=\rho^{(i)}_{\mathrm{G}}$ is practically not necessary.

\subsection{Setup of (II)}

In the second setup (II), the initial state is assumed to be a stationary state $\rho^{(i)}=\rho^{(i)}_{\mathrm{D}}$ (this assumption can be also removed under certain condition~\cite{Ikeda2015}, but we do not discuss it).
The final state is given by $\rho^{(f)}=e^{-iH_f\tau}\rho^{(i)}_{\mathrm{D}}e^{iH_f\tau}$ and the corresponding diagonal ensemble is given by
\beq
\rho^{(f)}_{\mathrm{D}}=\sum_n\<\phi^{(f)}_n|\rho^{(i)}_{\mathrm{D}}|\phi^{(f)}_n\>|\phi^{(f)}_n\>\<\phi^{(f)}_n|.
\eeq
The relative entropy $S(\rho^{(i)}_{\mathrm{D}}\|\rho^{(f)}_{\mathrm{D}})$ is then calculated as
\beq
S(\rho^{(i)}_{\mathrm{D}}\|\rho^{(f)}_{\mathrm{D}})=S(\rho^{(f)}_{\mathrm{D}})-S(\rho^{(i)}_{\mathrm{D}})
=S_{\mathrm{D}}(\rho^{(f)})-S_{\mathrm{D}}(\rho^{(i)}).
\eeq
From the non-negativity of the relative entropy, $S_{\mathrm{D}}(\rho^{(f)})-S_{\mathrm{D}}(\rho^{(i)})\geq 0$, which is the usual second law for the diagonal ensemble.

By using (\ref{eq:ineq_relative}) instead of the non-negativity of the relative entropy, we obtain
\beq
S_{\mathrm{D}}(\rho^{(f)})-S_{\mathrm{D}}(\rho^{(i)})\geq NI_f(x_i^*)+o(N),
\label{eq:ineq_diagonal}
\eeq
where $x_i^*$ is defined as the typical value of $X$ in the state $\rho^{(i)}_{\mathrm{D}}$ this time ($x_f^*$ is similarly defined), and $I_f(x)$ is a shorthand notation of $I_{\rho^{(f)}_{\mathrm{D}}}(x)$.

However, we cannot immediately conclude $S_{\mathrm{D}}(\rho^{(f)})-S_{\mathrm{D}}(\rho^{(i)})=\mathcal{O}(N)$ because $I_f(x_i^*)$ may be zero even for nontrivial quench.
As for the Gibbs state, it is reasonable to assume that the typical value of a macrovariable is unique~\cite{Netocny2004,Lenci2005,Ogata2010}, but the uniqueness of the typical value is not at all guaranteed for $\rho^{(f)}_{\mathrm{D}}$.
The requirement of nontrivial quench is merely the existence of $x_f^*\neq x_i^*$ with $I_f(x_f^*)=0$, and the possibility of $I_f(x_i^*)=0$ cannot be excluded.

We shall argue that under the assumption of eigenstate thermalization discussed below, we can exclude this possibility and conclude the strict irreversibility.
Now let us assume that both of $H_i$ and $H_f$ satisfy the eigenstate thermalization hypothesis (ETH).
The ETH states that every energy eigenstate has the property of thermal equilibrium.
There are several definitions of ETH reflected by several possible definitions of the equilibrium state~\cite{Neumann1929,Deutsch1991,Srednicki1994,Rigol2008}, but here we follow the definition provided in Ref.~\cite{Tasaki2016_typicality} because this version of the ETH is suited for the description of macroscopic properties in terms of macrovariables.

We consider the microcanonical energy shell $\mathcal{H}_{E,\Delta E}^{(a)}$ defined by
\beq
\mathcal{H}_{E,\Delta E}^{(a)}:=\mathrm{Span}\left\{|\phi_n^{(a)}\>: E-\Delta E\leq E_n^{(a)}\leq E+\Delta E\right\},
\eeq
where $a=i$ or $f$.
The microcanonical ensemble in this space is defined as
\beq
\rho_{\mathrm{mc}}^{(a)}:=\frac{1}{W}\sum_{|\phi_n^{(a)}\>\in\mathcal{H}_{E,\Delta E}^{(a)}}|\phi_n^{(a)}\>\<\phi_n^{(a)}|,
\eeq
where $W:=\sum_{|\phi_n^{(a)}\>\in\mathcal{H}_{E,\Delta E}^{(a)}}1$.
In the microcanonical ensemble, we assume that the typical value of a macrovariable is unique; $\mathcal{E}_{\rho_{\mathrm{mc}}^{(a)}}=\{x_{\rm eq}^{(a)}\}$.
This assumption corresponds to the ``thermodynamic bound'' discussed in Ref.~\cite{Tasaki2016_typicality}.

The ETH states that for every energy eigenstate $|\phi_n^{(a)}\>\in\mathcal{H}_{E,\Delta E}^{(a)}$ and for any fixed value of $\delta>0$, there exists $\gamma>0$ such that
\beq
\<\phi_n^{(a)}|\mathsf{P}_{|X-x_{\mathrm{eq}}^{(a)}|>\delta}|\phi_n^{(a)}\>\leq e^{-\gamma N}
\label{eq:ETH}
\eeq
for any macrovariable $X$ that satisfies the thermodynamic bound (or equivalently, the condition that the typical value in the microcanonical ensemble is unique).
Here, $\mathsf{P}_{|X-x_{\mathrm{eq}}^{(a)}|>\delta}$ is the projection operator onto the Hilbert subspace spanned by $\{|\psi\>: X|\psi\>=x|\psi\>, |x-x_{\mathrm{eq}}^{(a)}|>\delta\}$.

We now assume that the initial state belongs to $\mathcal{H}_{E_i,\Delta E}^{(i)}$ in the sense that
\beq
\rho^{(i)}=\rho^{(i)}_{\mathrm{D}}=\sum_{|\phi_n^{(i)}\>\in\mathcal{H}_{E_i,\Delta E}^{(i)}}p_n^{(i)}|\phi_n^{(i)}\>\<\phi_n^{(i)}|
\eeq
with $p_n^{(i)}\geq 0$ and $\sum_{|\phi_n^{(i)}\>\in\mathcal{H}_{E_i,\Delta E}^{(i)}}p_n^{(i)}=1$.

Considering the case of $X=H_f/N$, we have for an arbitrary fixed value of $\delta'>0$
\beq
\mathrm{Tr}\rho^{(i)}_{\mathrm{D}}\mathsf{P}_{|H_f-E_f|>\delta'N}\leq e^{-\gamma'N}
\label{eq:TBi}
\eeq
for some $\gamma'>0$ and $E_f=\mathrm{Tr}\rho_{\mathrm{mc}}^{(i)}H_f$.

Let us define the projection operator $\mathsf{P}:=\mathsf{P}_{|H_f-E_f|\leq\delta'N}$ and $\mathsf{Q}:=1-\mathsf{P}=\mathsf{P}_{|H_f-E_f|>\delta'N}$.
The diagonal ensemble after the quench is then decomposed as
\beq
\rho^{(f)}_{\mathrm{D}}=\mathsf{P}\rho^{(f)}_{\mathrm{D}}\mathsf{P}+\mathsf{Q}\rho^{(f)}_{\mathrm{D}}\mathsf{Q}.
\eeq
We obtain
\begin{align}
\mathrm{Tr}\mathsf{Q}\rho^{(f)}_{\mathrm{D}}\mathsf{Q}&=\mathrm{Tr}\mathsf{Q}\sum_n\<\phi_n^{(f)}|\rho^{(i)}_{\mathrm{D}}|\phi_n^{(f)}\>|\phi_n^{(f)}\>\<\phi_n^{(f)}|
\nonumber \\
&=\mathrm{Tr}\mathsf{Q}\rho^{(i)}_{\mathrm{D}}
\nonumber \\
&=\mathrm{Tr}\mathsf{P}_{|H_f-E_f|>\delta'N}\rho^{(i)}_{\mathrm{D}}\leq e^{-\gamma'N},
\end{align}
where we have used (\ref{eq:TBi}).
By using this inequality and (\ref{eq:ETH}), we obtain for any fixed value of $\delta>0$,
\begin{align}
\mathrm{Tr}\rho^{(f)}_{\mathrm{D}}\mathsf{P}_{|X-x_{\mathrm{eq}}^{(f)}|>\delta}
&=\mathrm{Tr}\mathsf{P}\rho^{(f)}_{\mathrm{D}}\mathsf{P}\mathsf{P}_{|X-x_{\mathrm{eq}}^{(f)}|>\delta}+\mathrm{Tr}\mathsf{Q}\rho^{(f)}_{\mathrm{D}}\mathsf{Q}\mathsf{P}_{|X-x_{\mathrm{eq}}^{(f)}|>\delta}
\nonumber \\
&\leq\sum_{|\phi_n^{(f)}\>\in\mathcal{H}_{E_f,N\delta}}\<\phi_n^{(f)}|\rho^{(f)}_{\mathrm{D}}|\phi_n^{(f)}\>\<\phi_n^{(f)}|\mathsf{P}_{|X-x_{\mathrm{eq}}^{(f)}|>\delta}|\phi_n^{(f)}\> +e^{-\gamma'N}
\nonumber \\
&\leq e^{-\gamma N}+e^{-\gamma'N}\leq 2e^{-\gamma''N},
\end{align}
where $\gamma'':=\min\{\gamma,\gamma'\}$.
This inequality implies that
\beq
I_f(x)\geq \gamma'' \text{ for all $x$ with $|x-x_{\mathrm{eq}}^{(f)}|>\delta$},
\eeq
and hence the typical value of a macrovariable $X$ is unique, $\mathcal{E}_{\rho^{(f)}_{\mathrm{D}}}=\{x_{\mathrm{eq}}^{(f)}\}$.

Let us go back to (\ref{eq:ineq_diagonal}).
It has been shown that if $H_i$ and $H_f$ satisfy the ETH in the sense of (\ref{eq:ETH}), we have, for any nontrivial quench with $x_i^*\neq x_f^*=x_{\mathrm{eq}}^{(f)}$ and $|x_f^*-x_i^*|>\delta$,
\beq
I_f(x_i^*)\geq\gamma''.
\eeq
Here, $\gamma''$ depends on $\delta$ but is independent of $N$.
From (\ref{eq:ineq_diagonal}), we obtain the desired inequality
\beq
S_{\mathrm{D}}(\rho^{(f)})-S_{\mathrm{D}}(\rho^{(i)})\geq N\gamma''+o(N)=\mathcal{O}(N).
\eeq
The diagonal entropy must also increase extensively.
This is the statement of the second law of thermodynamics with strict irreversibility for the diagonal entropy in quantum quench.

\section{Discussion}
\label{sec:summary}

In this paper, the second law of thermodynamics with strict irreversibility is proved in quantum quench for the thermodynamic entropy and the diagonal entropy.
For the thermodynamic entropy, the strict irreversibility is always true as long as the initial state is locally indistinguishable from the equilibrium state.
For the diagonal entropy, the strict irreversibility is rigorously derived when the initial and the final Hamiltonians obey the eigenstate thermalization hypothesis.

It should be emphasized that the extensive increase of the entropy in quantum quench is quite natural but far from trivial.
This is beyond the scope of thermodynamics; we cannot conclude the strict irreversibility only from the principle of thermodynamics.

There are remaining important future problems.
The first one is rather technical problem.
The quantum quench has been also extensively studied for integrable systems~\cite{Calabrese2007,Rigol2007,Calabrese2011,Vidmar_arXiv2016}, in which the ETH does not hold, triggered by experimental realization of integrable isolated quantum systems~\cite{Kinoshita2004,Kinoshita2006}.
In integrable systems, the increase of the diagonal entropy in quantum quench is numerically computed and the extensive increase of the diagonal entropy has been observed for several models~\cite{Santos2011,Vidmar_arXiv2016}.
It is an interesting problem to answer whether the extensive increase of the diagonal entropy is a general feature even for the systems without the ETH.

Second problem is to prove (or disprove) the extensive increase of the entropy for an arbitrary adiabatic process taking a finite time.
In order to somehow extend our result to an arbitrary non-quasi-static adiabatic process, new ideas will be necessary.
A remarkable recent progress is that it was shown that, in the setup of the classical Markov process, the Carnot efficiency in any finite-time cyclic process is impossible~\cite{Shiraishi2016}.
This is not a proof of the strict irreversibility, but would be a promising step towards solving this open problem.

This second future problem is particularly important and fascinating.
When we solve it, we will get profound understanding on the origin of irreversibility.

\section*{Acknowledgments}
The author would like to thank Tatsuhiko Shirai, Naoto Shiraishi, and Hal Tasaki for carefully reading the manuscript.
This work was financially supported by JSPS KAKENHI Grant No. 15K17718.

\end{document}